# Plateau-Rayleigh Instability of a Cylinder of Viscous Liquid
## (Rayleigh vs. Chandrasekhar)

L. Pekker

FujiFilm Dimatix Inc., Lebanon NH 03766 USA

Abstract

In 1892, in his classical work, L. Rayleigh considered the instability of a cylinder of viscous liquid under capillary force, the so-called Plateau-Rayleigh instability. In this work, in linear approximation, he obtained a dispersion equation describing the increment of this instability as a function of wavelength, the radius of cylinder, the mass density, surface tension, and viscosity of the liquid. Hundreds of authors referred to this work, but none of them used his dispersion equation in its complete form; they used only the asymptotic solutions of his equation for zero or infinitely large viscosities. A reason for this is, probably, that Rayleigh's writing is very difficult and his dispersion equation is quite complex. Then, in 1961, S. Chandrasekhar, in his monograph, also considered the stability of a viscous cylindrical jet and obtained his dispersion equation which is also quite complex and differs from the one obtained by Rayleigh. As in the case of Rayleigh's dispersion equation, other works use only the asymptotic solution of Chandrasekhar's equation that corresponds to the case where the viscosity is very large in comparison to inertia. In this paper, I demonstrate that Chandrasekhar's dispersion equation is equivalent to Rayleigh's and then simplify their dispersion equations to a form which can be easily solved numerically for arbitrary values of viscosity. I also present Mathematica code to calculate the maximum increment of the Plateau-Rayleigh instability for given parameters of the jet. To illustrate how the code works, I apply it to a cylindrical jet to estimate its breakup.

## I. Introduction

In 1879, in his first work [1], Rayleigh considers the instability of a cylinder of inviscid liquid due to capillary force (this instability then was called the Plateau-Rayleigh instability [2,3]) in linear approximation, where the relative deformation of the cylinder $\varepsilon_0$ is small:

$$R = R_0(1 + \varepsilon_0 e^{i\omega t - ikz}) \tag{1}$$

Here $z$ is measured parallel to the axis, $k$ is the wavelength, $\omega$ is the radial frequency, $t$ is time, $R$ is the radius of the jet, $R_0$ is the unperturbed equilibrium jet radius determined by $P_0 = \gamma/R_0$, where $P_0$ is the equilibrium pressure of the liquid and $\gamma$ is the surface tension. In this work, Rayleigh derives the dispersion equation connecting $k$ and $\omega$ which can be presented in the following non-dimensional form:

$$\tilde{\omega}^2 = (ikR_0)(1 - k^2 R_0^2) \frac{J_1(ikR_0)}{J_0(ikR_0)}, \tag{2}$$

where

$$\tilde{\omega} = \frac{\omega}{\sqrt{\gamma/(\rho(R_0)^3)}} \tag{3}$$

is non-dimensional radial frequency and $J_0$ and $J_1$ are the Bessel function of zeroth and first orders, respectively. This work is written in a very comprehendible way so that it is not very difficult to follow the author's logic and calculations. Then, in 1892, in his second paper [3], Rayleigh extends his work [1] to the case of a viscous cylindrical jet. In [3], he obtains the dispersion equation for an arbitrary value of viscosity that can be presented in the following non-dimensional form:

$$(1 - k^2 R_0^2) \frac{kR_0}{\tilde{\omega}} \left(\frac{\tilde{k}^2 - k^2}{k^2 + \tilde{k}^2}\right) J_0'(ikR_0) =$$

$$-2\tilde{\mu}(R_0 k)^2 \left( J_0''(ikR_0) - \frac{2k\tilde{k}}{(k^2 + \tilde{k}^2)} \frac{J_0'(ikR_0)}{J_0'(i\tilde{k}R_0)} J_0''(i\tilde{k}R_0) - \frac{\tilde{k}}{k}\left(\frac{\tilde{k}^2 - k^2}{k^2 + \tilde{k}^2}\right) \frac{J_0'(ikR_0)}{J_0'(i\tilde{k}R_0)} J_0(i\tilde{k}R_0) \right) +$$

$$\frac{\tilde{\omega}}{kR_0} \left( (ikR_0) J_0(ikR_0) - \frac{2k^2}{(k^2 + \tilde{k}^2)} \frac{J_0'(ikR_0)}{J_0'(i\tilde{k}R_0)} (i\tilde{k}R_0) J_0(i\tilde{k}R_0) \right), \tag{4}$$

where

$$\tilde{\mu} = \frac{\mu}{\sqrt{\gamma \cdot \rho \cdot R_0}} \quad , \tag{5}$$

$$\check{k}^2 = k^2 \left(1 + i\frac{\tilde{\omega}}{\tilde{\mu}(R_0 k)^2}\right) \quad , \tag{6}$$

and $J_0'(x)$ and is $J_0''(x)$ are the first and second derivatives of $J_0(x)$. Hundreds of authors referred to this work, but to the best of my knowledge, none of them used Rayleigh's dispersion equation in its complete form; they used only an asymptotic solution for the case of very large viscosity and very large wavelength, where $\mu \gg \sqrt{\gamma \rho R_0}$ and $kR_0 \ll 1$, see for example excellent review [4] and references therein:

$$\tilde{\omega} = -\frac{i}{6\tilde{\mu}} \quad . \tag{7}$$

A possible reason that the authors have not used Eq. (4), is because [3] is written rather densely, making it is very difficult to follow the Rayleigh's writing, logic and calculations.

In 1961, S. Chandrasekhar, in his monograph [5], also considers the Plateau-Rayleigh stability of a viscous cylindrical jet and obtains his dispersion equation which is also quite complex and differs from the one obtained by Rayleigh:

$$\frac{1}{\tilde{\mu}^2}\left(\frac{(\hat{k}R_0)I_1(\hat{k}R_0)}{I_0(\hat{k}R_0)}\right)\left(1 - (\hat{k}R_0)^2\right) = -(\hat{k}R_0)^4 + (\check{k}R_0)^4 +$$
$$2(\hat{k}R_0)^2\left((\hat{k}R_0)^2 + (\check{k}R_0)^2\right)\left(\frac{I_1'(\hat{k}R_0)}{I_0(\hat{k}R_0)}\right)\left[1 - \left(\frac{2\hat{k}\check{k}}{\hat{k}^2+\check{k}^2}\right)\left(\frac{I_1(\hat{k}R_0)}{I_1(\check{k}R_0)}\right)\left(\frac{I_1'(\check{k}R_0)}{I_1'(\hat{k}R_0)}\right)\right] \quad , \tag{8}$$

where

$$\check{k}^2 = \hat{k}^2\left(1 + \frac{\tilde{\sigma}}{\tilde{\mu}(R_0\hat{k})^2}\right) \quad , \tag{9}$$

$\tilde{\sigma}$ is non-dimensional increment of instability given by

$$\tilde{\sigma} = \frac{\sigma}{\sqrt{\gamma/(\rho(R_0)^3)}} \quad , \tag{10}$$

$I_0(x)$ and $I_1(x)$ are the modified Bessel functions of the first kind of the zeroth and first orders respectively, and $I_1'(x)$ is the first derivative of $I_1(x)$. In Chandrasekhar's formulation, the deformation of the cylinder is given by

$$R = R_0\left(1 + \varepsilon_0 e^{i\hat{k}z+\sigma t}\right) \quad , \tag{11}$$

which differs from the form that Rayleigh used, see Eq. (1). Comparing Eqs. (1) and (11), one can see that the pairs $(k, \omega)$ and $(\hat{k}, \sigma)$ are connected by the following relationship

$$\hat{k} = -k \qquad \sigma = i\omega \quad . \tag{12}$$

The case where the viscosity is very large in comparison with inertia corresponds to $\left|\frac{\omega\rho}{\mu k}\right| \ll 1$ in the Rayleigh formulation, and to $\left|\frac{\sigma\rho}{\mu \hat{k}}\right| \ll 1$ in Chandrasekhar's formulation. In this case, the authors reduced their dispersion equations to the forms [3, 5] which can be presented in the following non-dimensional forms:

$$i\widetilde{\omega} = -\left(\frac{1}{2\widetilde{\mu}}\right)\frac{\left(1-(kR_0)^2\right)}{1+(kR_0)^2\left(1+\left(\frac{J_0(ikR_0)}{J_0'(ikR_0)}\right)^2\right)} \tag{13}$$

and

$$\widetilde{\sigma} = -\left(\frac{1}{2\widetilde{\mu}}\right)\frac{\left(1-(\hat{k}R_0)^2\right)}{1+(kR_0)^2\left(1-\left(\frac{I_0(\hat{k}R_0)}{I_1(\hat{k}R_0)}\right)^2\right)} \quad , \tag{14}$$

where Eq. (13) was obtained by Rayleigh [3] and Eq. (14) by Chandrasekhar [5]. Taking into account Eq. (12) and the following Bessel functions properties:

$$2J_n'(x) = J_{n-1}(x) - J_{n+1}(x) \quad J_n(-x) = J_{-n}(x) = (-1)^n J_n(x) \quad J_n(ix) = (i)^n I_n(x) \tag{15}$$

it is easy to see that Eqs. (13) and (14) are equivalent. As has been mentioned by Chandrasekhar in [5], in the case of inviscid liquid, his dispersion equations reduce to Rayleigh's [1]. Thus, the question is: are the dispersion equations obtained by Rayleigh and Chandrasekhar, Eqs. (4) and (8), equivalent for arbitrary values of $\mu$?

In section II, I demonstrate the step-by-step derivation of a new dispersion equation of the Plateau-Rayleigh instability for a cylindrical viscous jet which can be easily solved numerically. In section III, I demonstrate that both Rayleigh's and Chandrasekhar dispersion equations, Eqs. (4) and (8), can be reduced to the this new form and, therefore, all three dispersion equations are equivalent. In Section IV, I present Mathematica code that calculates the maximum increment of the Plateau-Rayleigh instability for given parameters of the jet: the radius of the jet and the mass density, the surface tension, and the viscosity of the liquid, and then demonstrate how the code works by applying it to a cylindrical jet to estimate its breakup. Conclusions are given in Section V.

## II. Derivation the dispersion equation for Plateau-Rayleigh instability

In this section, basically following the Rayleigh's path [3], I derive step-by-step the dispersion equation of the Plateau-Rayleigh instability for the case of an infinite cylinder of a viscous incompressible liquid. In linear approximation, the set of equations describing the Plateau-Rayleigh instability in cylindrical coordinates can be written as

$$\frac{\partial v_r}{\partial t} = -\frac{1}{\rho}\frac{\partial P}{\partial r} + \frac{\mu}{\rho}\left(\frac{1}{r}\frac{\partial}{\partial r}\left(r\frac{\partial v_r}{\partial r}\right) - \frac{v_r}{r^2} + \frac{\partial^2 v_r}{\partial z^2}\right), \tag{16}$$

$$\frac{\partial v_z}{\partial t} = -\frac{1}{\rho}\frac{\partial P}{\partial z} + \frac{\mu}{\rho}\left(\frac{1}{r}\frac{\partial}{\partial r}\left(r\frac{\partial v_z}{\partial r}\right) + \frac{\partial^2 v_z}{\partial z^2}\right), \tag{17}$$

$$\frac{\partial v_z}{\partial z} + \frac{1}{r}\frac{\partial}{\partial r}(rv_r) = 0, \tag{18}$$

$$R(z,t) = R_0\left(1 + \varepsilon_0 e^{i\omega t - ikz}\right) \text{ where } \varepsilon_0 \ll 1, \tag{19}$$

$$P(r,z,t) = P_0 + \delta P(r)e^{i\omega t - ikz}, \tag{20}$$

$$P_0 = \frac{\gamma}{R_0}, \tag{21}$$

where, Eqs. (16) – (17) are the $r$- and $z$- components of the Navier-Stokes equation; Eq. (18) is the mass conservation equation; Eq. (19) describes the deformation of the jet surface, see Eq. (1); Eq. (20) is an equation for the pressure where $P_0$, Eq. (21), is the equilibrium pressure of unperturbed jet.

A set of boundary conditions for these equations are:

$$\delta P(R_0) = \frac{\gamma\varepsilon_0}{R_0}\left(k^2 R_0^2 - 1\right) + 2\mu\left(\frac{\partial v_r}{\partial r}\right)_{r=R_0} e^{-(i\omega t - ikz)}, \tag{22}$$

$$\left(\frac{\partial v_r}{\partial z} + \frac{\partial v_z}{\partial r}\right)_{r=R_0} = 0 \quad , \tag{23}$$

$$\frac{\partial R}{\partial t} = (v_r)_{r=R_0} \quad , \tag{24}$$

$$v_r(r=0) = 0 \quad , \tag{25}$$

where Eqs. (22) – (23) correspond to the boundary conditions for the normal and tangential stresses at the free surface; Eq. (24) describes the movement of the liquid at the jet surface; and Eq. (25) states that radial velocity of the liquid at $r = 0$ is equal to zero. Eq. (22) - (24) are obtained in Appendix A.

Following [3], let us introduce the Stokes' stream function $\psi$ defined by

$$v_r = \frac{1}{r}\frac{\partial \psi}{\partial z} \qquad v_z = -\frac{1}{r}\frac{\partial \psi}{\partial r} \quad . \tag{26}$$

Substitution of Eq. (26) into Eq. (18) yields that Eq. (18) is automatically satisfied. Substituting Eq. (26) into Eqs. (16) and (17) yields

$$\frac{\partial}{\partial t}\left(\frac{1}{r}\frac{\partial \psi}{\partial z}\right) = -\frac{1}{\rho}\frac{\partial P}{\partial r} + \frac{\mu}{\rho}\frac{1}{r}\left(\frac{\partial^3 \psi}{\partial r^2 \partial z} - \frac{1}{r}\frac{\partial^2 \psi}{\partial r \partial z} + \frac{\partial^3 \psi}{\partial z^3}\right) \quad , \tag{27}$$

$$-\frac{\partial}{\partial t}\left(\frac{1}{r}\frac{\partial \psi}{\partial r}\right) = -\frac{1}{\rho}\frac{\partial P}{\partial z} - \frac{\mu}{\rho}\left(\frac{1}{r^3}\frac{\partial \psi}{\partial r} - \frac{1}{r^2}\frac{\partial^2 \psi}{\partial r^2} + \frac{1}{r}\frac{\partial^3 \psi}{\partial r^3}\right) - \frac{\mu}{\rho}\frac{1}{r}\frac{\partial^3 \psi}{\partial r \partial z^2} \quad . \tag{28}$$

Taking $z$-derivative of Eq. (27) and $r$-derivative of Eq. (28) and then eliminating pressure by subtracting one equation from the another, after simple algebra, I obtain

$$\frac{\rho}{\mu}\left(\frac{\partial^3 \psi}{\partial t \partial z^2} - \frac{1}{r}\frac{\partial^2 \psi}{\partial t \partial r} + \frac{\partial^3 \psi}{\partial t \partial r^2}\right) = 2\frac{\partial^4 \psi}{\partial r^2 \partial z^2} - 2\frac{1}{r}\frac{\partial^2 \psi}{\partial r \partial z^2} + \frac{\partial^4 \psi}{\partial z^4} - \frac{3}{r^3}\frac{\partial \psi}{\partial r} + \frac{3}{r^2}\frac{\partial^2 \psi}{\partial r^2} - 2\frac{1}{r}\frac{\partial^3 \psi}{\partial r^3} + \frac{\partial^4 \psi}{\partial r^4} \quad . \tag{29}$$

In [3], Eq. (29) is presented in the following form:

$$\left(\frac{\partial^2}{\partial r^2} - \frac{1}{r}\frac{\partial}{\partial r} + \frac{\partial^2}{\partial z^2} - \frac{\rho}{\mu}\frac{\partial}{\partial t}\right)\left(\frac{\partial^2}{\partial r^2} - \frac{1}{r}\frac{\partial}{\partial r} + \frac{\partial^2}{\partial z^2}\right)\psi = 0 \quad . \tag{30}$$

Simple calculation shows that Eq. (30) indeed can be reduced to Eq. (29). Following [3], let us present a solution of Eq. (30) (or Eq. (29)) as

$$\psi = e^{i(\omega t - kz)}\Phi_1(r) + e^{i(\omega t - kz)}\Phi_2(r), \tag{31}$$

where $\Phi_1(r)$ and $\Phi_2(r)$ are solutions of the following equations:

$$\frac{d^2\Phi_1}{dr^2} - \frac{1}{r}\frac{d\Phi_1}{dr} - k^2\Phi_1 = 0 \quad , \tag{32}$$

$$\frac{d^2\Phi_2}{dr^2} - \frac{1}{r}\frac{d\Phi_2}{dr} - \tilde{k}^2\Phi_2 = 0 \quad , \tag{33}$$

where

$$\tilde{k}^2 = \left(k^2 + i\frac{\rho\omega}{\mu}\right) \quad . \tag{34}$$

General solutions of Eqs. (32) and (33) can be presented in the following form:

$$\Phi_1(r) = Aikr J_1(ikr) \qquad \Phi_2(r) = \tilde{A} i\tilde{k} r J_1(i\tilde{k}r), \tag{35}$$

where $A$ and $\tilde{A}$ are arbitrary constantans, and $J_1$ is the Bessel function of the first order; derivation of Eqs. (35) is presented in Appendix B.

Substituting Eq. (31) along with Eqs. (35) into Eqs. (26), I obtain

$$v_r = \left(Ak^2 J_1(ikr) + \tilde{A}\tilde{k}k J_1(i\tilde{k}r)\right) e^{i\omega t - ikz} \quad , \tag{36}$$

$$v_z = \left(Ak^2 J_0(ikr) + \tilde{A}\tilde{k}^2 J_0(i\tilde{k}r)\right) e^{i\omega t - ikz} \quad . \tag{37}$$

In Eq. (37), I have taken into account that $d(xJ_1(x))/dx = xJ_0(x)$. As one can see from Eq. (36), the boundary condition $v_r(r = 0) = 0$, Eq. (25), is automatically satisfied.

Substituting Eqs. (36) and (37) into Eq. (17) and using the Bessel function properties,

$$\frac{d(x^n J_n(x))}{dx} = x^n J_{n-1}(x) \quad J_{-n}(x) = (-1)^n J_n(x) \quad , \tag{38}$$

I obtain an equation for $\delta P$:

$$i\omega \left(Ak^2 J_0(ikr) + \tilde{A}\tilde{k}^2 J_0(i\tilde{k}r)\right) - \frac{1}{\rho} ik\delta P(r) =$$

$$\frac{\mu}{\rho}\left(\frac{1}{r}\frac{\partial}{\partial r}\left(r\frac{\partial\left(Ak^2 J_0(ikr)+\tilde{A}\tilde{k}^2 J_0(i\tilde{k}r)\right)}{\partial r}\right)-\left(Ak^4 J_0(ikr)+\tilde{A}\tilde{k}^2 k^2 J_0(i\tilde{k}r)\right)\right)=$$

$$\frac{\mu}{\rho}\left(\tilde{k}^2\tilde{A}(\tilde{k}^2-k^2)J_0(i\tilde{k}r)\right)=i\omega\tilde{k}^2\tilde{A}J_0(i\tilde{k}r)\ \rightarrow$$

$$\delta P(r)=\rho\omega k A J_0(ikr) \tag{39}$$

In Eq. (39), I have taken into account that $\tilde{k}^2-k^2=i\rho\omega/\mu$, Eq. (34).

## IIa. Boundary condition Eq. (22)

Substituting $\delta P(r)$ from Eq. (39) and $v_r$ from Eq. (36) into Eq. (22), I obtain:

$$\rho\omega k A J_0(ikR_0)=\frac{\gamma\varepsilon_0}{R_0}\left(k^2 R_0^2-1\right)+2\mu\left(Ak^2\frac{\partial J_1(ikr)}{\partial r}+\tilde{A}\tilde{k}k\frac{\partial J_1(i\tilde{k}r)}{\partial r}\right)_{r=R_0}. \tag{40}$$

Using Eq. (15) reduces Eq. (40) to the following form

$$A\rho\omega k J_0(ikR_0)=$$
$$=\frac{\gamma\varepsilon_0}{R_0}((kR_0)^2-1)+\mu ik\left(Ak^2(J_0(ikR_0)-J_2(ikR_0))+\tilde{A}\tilde{k}^2\left(J_0(i\tilde{k}R_0)-J_2(i\tilde{k}R_0)\right)\right). \tag{41}$$

## IIb. Boundary condition Eq. (23)

Substituting Eqs. (36) and (37) into Eq. (23) gives

$$-ik\left(Ak^2 J_1(ikr)+\tilde{A}\tilde{k}k J_1(i\tilde{k}r)\right)_{r=R_0}+\left(k^2 A\frac{dJ_0(ikr)}{dr}+\tilde{k}^2\tilde{A}\frac{dJ_0(i\tilde{k}r)}{dr}\right)_{r=R_0}\rightarrow$$

$$2Ak^3 J_1(ikR_0)+(k^2+\tilde{k}^2)\tilde{k}\tilde{A}J_1(i\tilde{k}R_0)=0\ \rightarrow$$

$$\tilde{A}=-\frac{2k^3 J_1(ikR_0)}{(k^2+\tilde{k}^2)\tilde{k}J_1(i\tilde{k}R_0)}A\ . \tag{42}$$

In Eq. (42), I have used Eq. (38).

### IIc. Boundary condition Eq. (24)

Substituting Eqs. (19) and (36) into Eq. (24), I obtain

$$i\omega R_0 \varepsilon_0 = \left(Ak^2 J_1(ikR_0) + \tilde{A}\tilde{k}k J_1(i\tilde{k}R_0)\right) \quad . \tag{43}$$

### IId. Dispersion Equation

The system of Eqs. (41), (42), and (43) for $\varepsilon_0$, $A$, and $\tilde{A}$ along with Eq. (34) for $\tilde{k}$ determine the dispersion equation connecting $k$ and $\omega$. Substituting $\varepsilon_0$ from Eq. (43) into Eq. (41), I obtain

$$\frac{\gamma}{\omega R_0^3}(1-(kR_0)^2)\left(AikR_0 J_1(ikR_0) + \tilde{A}i\tilde{k}R_0 J_1(i\tilde{k}R_0)\right) =$$

$$-\mu i \left(Ak^2 (J_0(ikR_0) - J_2(ikR_0)) + \tilde{A}\tilde{k}^2 \left(J_0(i\tilde{k}R_0) - J_2(i\tilde{k}R_0)\right)\right) + A\rho\omega J_0(ikR_0) \quad . \tag{44}$$

Substituting $\tilde{A}$ from Eq. (42) into Eq. (44) and then using Eqs. (3) and (5), I obtain the following dispersion equation for Plateau-Rayleigh instability in a non-dimensional form:

$$(1-(kR_0)^2)ikR_0 \frac{J_1(ikR_0)}{J_0(ikR_0)} \cdot \left(\frac{\tilde{k}^2 - k^2}{k^2 + \tilde{k}^2}\right) = \tilde{\omega}^2 -$$

$$i(kR_0)^2 \tilde{\omega}\tilde{\mu}\left(1 - \frac{J_2(ikR_0)}{J_0(ikR_0)} - \frac{2k\tilde{k}}{(k^2+\tilde{k}^2)} \cdot \frac{J_1(ikR_0)}{J_1(i\tilde{k}R_0)} \left(\frac{J_0(i\tilde{k}R_0) - J_2(i\tilde{k}R_0)}{J_0(ikR_0)}\right)\right) \quad . \tag{45}$$

Introducing $\xi = kR_0$ and using $\tilde{k}^2$ in a form given by Eq. (6), I obtain:

$$\left(\frac{\tilde{k}^2 - k^2}{k^2 + \tilde{k}^2}\right) = \left(1 - 2i\frac{\tilde{\mu}\xi^2}{\tilde{\omega}}\right)^{-1} \quad \frac{2k\tilde{k}}{(k^2+\tilde{k}^2)} = \frac{\left(1+i\frac{\tilde{\omega}}{\tilde{\mu}\xi^2}\right)^{1/2}}{\left(1+i\frac{\tilde{\omega}}{2\tilde{\mu}\xi^2}\right)} \quad \tilde{k}R_0 = \xi\left(1 + i\frac{\tilde{\omega}}{\tilde{\mu}\xi^2}\right)^{1/2} \quad . \tag{46}$$

Substituting Eqs. (46) into Eq. (45) I reduce the dispersion Eq. (45) to the following form:

$$(1-\xi^2)i\xi \frac{J_1(i\xi)}{J_0(i\xi)} = \tilde{\omega}^2 - 2i\tilde{\mu}\tilde{\omega}\xi^2 - i\tilde{\mu}\xi^2(\tilde{\omega} - 2i\tilde{\mu}\xi^2) \cdot$$

$$\left(1 - \frac{J_2(i\xi)}{J_0(i\xi)} - \frac{\left(1+i\frac{\widetilde{\omega}}{\widetilde{\mu}\xi^2}\right)^{1/2}}{\left(1+i\frac{\widetilde{\omega}}{2\widetilde{\mu}\xi^2}\right)} \frac{J_1(i\xi)}{J_1\left(i\xi\left(1+i\frac{\widetilde{\omega}}{\widetilde{\mu}\xi^2}\right)^{1/2}\right)} \left(\frac{J_0\left(i\xi\left(1+i\frac{\widetilde{\omega}}{\widetilde{\mu}\xi^2}\right)^{1/2}\right) - J_2\left(i\xi\left(1+i\frac{\widetilde{\omega}}{\widetilde{\mu}\xi^2}\right)^{1/2}\right)}{J_0(i\xi)}\right)\right) \quad . \quad (47)$$

Substituting $k = -\hat{k}$ and $\sigma = i\omega$, Eq. (12), into Eq. (47) and introducing $\hat{\xi} = -\xi$, I obtain the dispersion equation in terms of Chandrasekhar notation, Eq. (11):

$$(1 - \hat{\xi}^2)\hat{\xi}\frac{I_1(\hat{\xi})}{I_0(\hat{\xi})} = \tilde{\sigma}^2 + 2\tilde{\mu}\tilde{\sigma}\hat{\xi}^2 + \tilde{\mu}\hat{\xi}^2(\tilde{\sigma} + 2\tilde{\mu}\hat{\xi}^2) \cdot$$

$$\left(1 + \frac{I_2(\hat{\xi})}{I_0(\hat{\xi})} - \frac{\left(1+\frac{\tilde{\sigma}}{\tilde{\mu}\hat{\xi}^2}\right)^{1/2}}{\left(1+\frac{\tilde{\sigma}}{2\tilde{\mu}\hat{\xi}^2}\right)} \frac{I_1(\hat{\xi})}{I_1\left(\hat{\xi}\left(1+\frac{\tilde{\sigma}}{\tilde{\mu}\hat{\xi}^2}\right)^{1/2}\right)} \left(\frac{I_0\left(\hat{\xi}\left(1+\frac{\tilde{\sigma}}{\tilde{\mu}\hat{\xi}^2}\right)^{1/2}\right) + I_2\left(\hat{\xi}\left(1+\frac{\tilde{\sigma}}{\tilde{\mu}\hat{\xi}^2}\right)^{1/2}\right)}{I_0(\hat{\xi})}\right)\right) \quad . \quad (48)$$

In deriving Eq. (48) I used Eqs. (15) and the following Bessel function properties

$$2I'_n(x) = I_{n-1}(x) + I_{n+1}(x) \quad I_{-n}(x) = I_n(x) = (-1)^n J_n(-x) \quad . \quad (49)$$

It is worth noting that in the case of inviscid liquid ($\tilde{\mu} = 0$), Eqs. (47) and (48) reduce to Eq. (2).

## III. Rayleigh's and Chandrasekhar's dispersion equations

In this Section, I will demonstrate that Rayleigh's dispersion Eq. (4) and Chandrasekhar's dispersion Eq. (8) are equivalent by reducing them to dispersion Eqs. (45) and (47) respectively.

Let us first consider the case of Rayleigh's dispersion equation. Multiplying Eq. (4) by the factor $-i\widetilde{\omega}/J_0(ikR_0)$ and then using the relationship $J''_0(ikR_0) = -0.5\big(J_0(ikR_0) - J_2(ikR_0)\big)$ that also follows from Eq. (15) as well as simple algebra, I obtain

$$(1 - k^2 R_0^2) ikR_0 \left(\frac{\tilde{k}^2 - k^2}{k^2 + \tilde{k}^2}\right) \frac{J_1(ikR_0)}{J_0(ikR_0)} =$$

$$= \widetilde{\omega}^2 - i\tilde{\mu}\widetilde{\omega}(R_0 k)^2 \left(1 - \frac{J_2(ikR_0)}{J_0(ikR_0)} - \frac{2k\tilde{k}}{(k^2+\tilde{k}^2)} \frac{J_1(ikR_0)}{J_1(i\tilde{k}R_0)} \left(\frac{J_0(i\tilde{k}R_0) - J_2(i\tilde{k}R_0)}{J_0(ikR_0)}\right)\right)$$

$$-i\tilde{\mu}\widetilde{\omega}(R_0)^2 \left(\frac{2\tilde{k}k}{k^2+\tilde{k}^2} \frac{J_1(ikR_0)}{J_1(i\tilde{k}R_0)} \frac{J_0(i\tilde{k}R_0)}{J_0(ikR_0)}\right)(\tilde{k}^2 - k^2) - \widetilde{\omega}^2 \frac{2k\tilde{k}}{(k^2+\tilde{k}^2)} \frac{J'_0(i\tilde{k}R_0)}{J'_0(ikR_0)} \frac{J_0(i\tilde{k}R_0)}{J_0(ikR_0)} \quad .$$

Substituting $\tilde{k}^2 - k^2$ from Eq. (6) into the first term on the last line of this equation, this term cancels the next term on this line, thereby yielding Eq. (45).

Now I will consider the case of Chandrasekhar's dispersion equation, Eq. (8). Introducing $\hat{\xi}$ into Eq. (8) and then substituting $\check{k}^2$ from Eq. (9), Eq. (8) reduces to the following form:

$$\frac{1}{\tilde{\mu}^2}(1-\xi^2)\xi\frac{I_1(\hat{\xi})}{I_0(\hat{\xi})} = -\xi^4 + \left(\hat{\xi}^2 + \frac{\tilde{\sigma}}{\tilde{\mu}}\right)^2 +$$

$$2(\hat{\xi})^2\left(\hat{\xi}^2 + \left(\hat{\xi}^2 + \frac{\tilde{\sigma}}{\tilde{\mu}}\right)\right)\left(\frac{I'_1(\hat{\xi})}{I_0(\hat{\xi})}\right)\left[1 - \left(\frac{\left(1+\frac{\tilde{\sigma}}{\tilde{\mu}\xi^2}\right)^{1/2}}{\left(1+\frac{\tilde{\sigma}}{2\tilde{\mu}\xi^2}\right)}\right)\left(\frac{I_1(\hat{\xi})}{I_1\left(\hat{\xi}\left(1+\frac{\tilde{\sigma}}{\tilde{\mu}\xi^2}\right)^{1/2}\right)}\right)\left(\frac{I'_1\left(\hat{\xi}\left(1+\frac{\tilde{\sigma}}{\tilde{\mu}\xi^2}\right)^{1/2}\right)}{I'_1(\hat{\xi})}\right)\right],$$

Multiplying this equation by $\tilde{\mu}^2$ and using Eqs. (49), this equation can be reduced to the following form

$$(1-\xi^2)\xi\frac{I_1(\hat{\xi})}{I_0(\hat{\xi})} = \tilde{\sigma}^2 + 2\hat{\xi}^2\tilde{\mu}\tilde{\sigma} + \tilde{\mu}\xi^2(\tilde{\sigma} + 2\hat{\xi}^2\tilde{\mu}) \cdot$$

$$\cdot \left(\frac{I_0(\hat{\xi})+I_2(x)}{I_0(\hat{\xi})}\right)\left[1 - \frac{\left(1+\frac{\tilde{\sigma}}{\tilde{\mu}\xi^2}\right)^{1/2}}{\left(1+\frac{\tilde{\sigma}}{2\tilde{\mu}\xi^2}\right)}\frac{I_1(\hat{\xi})}{I_1\left(\hat{\xi}\left(1+\frac{\tilde{\sigma}}{\tilde{\mu}\xi^2}\right)^{1/2}\right)}\left(\frac{I_0\left(\hat{\xi}\left(1+\frac{\tilde{\sigma}}{\tilde{\mu}\xi^2}\right)^{1/2}\right)+I_2\left(\hat{\xi}\left(1+\frac{\tilde{\sigma}}{\tilde{\mu}\xi^2}\right)^{1/2}\right)}{I_0(\hat{\xi})+I_2(x)}\right)\right].$$

Bringing term $(I_0(\hat{\xi}) + I_1(x))/I_0(\hat{\xi})$ inside of the square brackets, I obtain Eq. (48).

Thus, I have shown that Rayleigh's and Chandrasekhar's dispersion equations are equivalent and can be reduced to the forms given by Eq. (45) (or Eq. (47)) and (48) respectively.

## IV. Numerical results

The following Mathematica code solves Eq. (47) to determine the maximum increment of the Plateau-Rayleigh instability for cylindrical viscous liquid jet:

```
μhat=50.0;
LFS[ξ_]:=(1-ξ^2)*I*ξ*(BesselJ[1,I*ξ]/BesselJ[0,I*ξ]);
RHS1[ξ_, ωhat_] := ωhat^2 - 2*I*μhat*ωhat*ξ^2;
RHS2a[ξ_, ωhat_] := -I*μhat*ξ^2*(ωhat - 2*I*μhat*ξ^2);
RHS2b[ξ_, ωhat_] := (1 - BesselJ[2, I*ξ]/BesselJ[0, I*ξ]) - ((1 + I*ωhat /( μhat*ξ^2))^0.5/
      (1 + I*ωhat/(2* μhat*ξ^2))*( BesselJ[1, I*ξ]/
      BesselJ[1, I*ξ*((1 + I*ωhat/(μhat*ξ^2))^0.5)])*((BesselJ[0, I*ξ*((1 + I*ωhat/(μhat*ξ^2))^0.5)]
```

```
             - BesselJ[2, I*ξ*((1 + I*ωhat/(μhat*ξ^2))^0.5)])/BesselJ[0, I*ξ]);
RHSA[ξ_, ωhat_] := RHS1[ξ, ωhat] + RHS2a[ξ, ωhat]*RHS2b[ξ, ωhat];
ωhatD[ξ_] := ωhat /. FindRoot[LFS[ξ] == RHSA[ξ, ωhat], {ωhat, -I-1}];
ImωhatDmax=MaxValue[{-Im[ωhatD[ξ]],ξ>0,ξ<1.5},ξ];
ξD:=ξ/.FindRoot[-Im[ωhatD[ξ]]==ImωhatDmax,{ξ,1}];
Print["ξD = ",ξD," -ImωhatAmax = ",ImωhatDmax];
    ξD =  0.0683365   -ImωhatAmax =  0.00330219
```

In this code, $\mu hat$, $\omega hat$, and $\xi$ correspond to $\tilde{\mu}$, $\tilde{\omega}$, and $\xi = kR_0$ respectively; $\xi D$ corresponds to $\xi$ at which the increment of the instability, $-i\tilde{\omega}$, reaches its maximal value $-i\tilde{\omega}_{max}$; and $Im\omega hatAmax$ corresponds to $-i\tilde{\omega}_{max}$. Using this code I have calculated the maximum increment and the corresponding $\xi_{max}$ as functions on $\tilde{\mu}$ that are presented Figs. 1 and 2. The characteristic non-dimensional time of the development of the Plateau-Rayleigh instability can be written as

$$\tilde{\tau} = \frac{1}{-Im(\tilde{\omega})} = \tau \left(\frac{\gamma}{\rho(R_0)^3}\right)^{1/2} \quad , \tag{50}$$

where $\tau$ is the increment of the instability in seconds. As one can see from Fig. 3, with good accuracy, $\tilde{\tau}_{min}$ can be approximated as

$$\tilde{\tau}_{min} = 5.9982\tilde{\mu} + 2.8778 \quad . \tag{51}$$

Substituting Eqs. (50) and (5) into Eq. (51), I obtain that the characteristic time of the Plateau-Rayleigh instability jet can be approximated as

$$\tau_{min} = 5.9982\frac{\mu R_0}{\gamma} + 2.8778\left(\frac{\gamma}{\rho(R_0)^3}\right)^{1/2} \quad . \tag{52}$$

The non-dimensional wavelength which corresponds to $\tilde{\tau}_{min}$, $\tilde{\lambda}_0 = 2\pi/\xi_{max}$, is presented in Fig. 4. As one can see, for $\tilde{\mu} < 1$, with good accuracy $\tilde{\lambda}_0$ can be approximated as

$$\tilde{\lambda}_0 = 6.9345\tilde{\mu} + 9.1174 \quad \tilde{\mu} < 1 \quad , \tag{53}$$

Substituting $\tilde{\mu}$ from Eq. (5) into Eq. (53) and taking into account that $\lambda = \tilde{\lambda}R_0$, I obtain that

$$\lambda_0 = 6.9345\mu\sqrt{\frac{R_0}{\gamma\rho}} + 9.1174 R_0 \qquad \frac{\mu}{\sqrt{\gamma\rho R_0}} < 1 \qquad . \qquad (54)$$

Eqs. (52) and (54) can be used to estimate a characteristic time and characteristic length of breakdown of the jet.

## V. Conclusions

In this paper, I have demonstrated that Chandrasekhar's dispersion equation [5] for the Plateau-Rayleigh instability for a cylinder of viscous incompressible liquid is equivalent to the dispersion equation obtained by Rayleigh [1]. I reduced these dispersion equations to an equation which can be easily solved numerically. A Mathematica code solving this equation is presented and applied to estimate of liquid jet breakdown.

## Acknowledgments

I would like to express my sincere gratitude to my colleagues Matthew Aubrey who encouraged me to read the Rayleigh's papers and kindly supported me during this research and also James Myrick, Paul Hoisington, and Dan Barnett for their valuable comments. I also would like to thank Alexander Pekker for his kind help in preparation the text of this paper.

## Appendix A

The stress tensor of the axisymmetric incompressible jet in the cylindrical system of coordinates is

$$\vec{\vec{S}} = \mu \begin{pmatrix} 2\frac{\partial v_r}{\partial r} & 0 & \frac{\partial v_z}{\partial r} + \frac{\partial v_r}{\partial z} \\ 0 & 2\frac{v_r}{r} & 0 \\ \frac{\partial v_z}{\partial r} + \frac{\partial v_r}{\partial z} & 0 & 2\frac{\partial V_z}{\partial z} \end{pmatrix}_{r=R} \qquad (A1)$$

and the normal and tangential vectors at the surface of the jet are

$$\vec{n} = \frac{\left(1,0,-\frac{\partial R}{\partial z}\right)}{\sqrt{\left(\frac{\partial R}{\partial z}\right)^2 + 1}} \quad \text{and} \quad \vec{\tau} = \frac{\left(\frac{\partial R}{\partial z},0,1\right)}{\sqrt{\left(\frac{\partial R}{\partial z}\right)^2 + 1}} \qquad , \qquad (A2)$$

where $R = R(z,t)$ is the radius of the jet. The first boundary condition at the free surface (Eq. (23)) is that tangential stress at the free surface is equal to zero, which yields:

$$\vec{n} \cdot \overleftrightarrow{S} \cdot \vec{\tau} = 0 \rightarrow \left(1 - \left(\frac{\partial R}{\partial z}\right)^2\right)\left(\frac{\partial v_z}{\partial r} + \frac{\partial v_r}{\partial z}\right)_{r=R} + 2\left(\frac{\partial v_r}{\partial r} - \frac{\partial v_z}{\partial z}\right)_{r=R} \frac{\partial R}{\partial z} = 0 \qquad (A3)$$

Substituting Eq. (19) into Eq. (A3) and neglecting the terms of the first order in $\varepsilon_0$ yields

$$\left(\frac{\partial v_r}{\partial z} + \frac{\partial v_z}{\partial r}\right)_{r=R} = 0 \qquad . \qquad (A4)$$

The second boundary condition at free surface (Eq. (22)) is that the normal stress at the free surface is balanced by the external pressure and the surface tension, which yields:

$$P - \vec{n} \cdot \overleftrightarrow{S} \cdot \vec{n} = \gamma\left(\frac{1}{R_1} + \frac{1}{R_2}\right) \rightarrow$$

$$P - \frac{2\mu}{\left(\frac{\partial R}{\partial z}\right)^2 + 1}\left(\left(\frac{\partial v_r}{\partial r}\right)_{r=R} - \frac{\partial R}{\partial z}\left(\frac{\partial v_z}{\partial r} + \frac{\partial v_r}{\partial z}\right)_{r=R} + \left(\frac{\partial R}{\partial z}\right)^2 \left(\frac{\partial v_z}{\partial z}\right)_{r=R}\right) = \gamma\left(\frac{1}{R_1} + \frac{1}{R_2}\right) , \qquad (A5)$$

where

$$R_1 = R\left(1 + \left(\frac{\partial R}{\partial z}\right)^2\right)^{1/2} \qquad R_2 = -\left(\frac{\partial^2 R}{\partial z^2}\right)^{-1} \cdot \left(1 + \left(\frac{\partial R}{\partial z}\right)^2\right)^{3/2} \qquad (A6)$$

are the curvature radii of the jet surface. Substituting $R$ from Eqs. (19) into Eqs. (A.6) and then neglecting all terms which are of the order $(\varepsilon_0)^2$, I obtain that

$$\frac{1}{R_1} = \frac{1}{R_0} - \frac{\varepsilon_0}{R_0} e^{i\omega t - ikz} \qquad \frac{1}{R_2} = R_0 \varepsilon_0 k^2 e^{i\omega t - ikz} \qquad . \qquad (A7)$$

Substituting Eqs. (A7), (20), and (21) into Eqs. (A5) and then taking into account Eq. (A4) and neglecting all terms of the order $(\varepsilon_0)^2$, I obtain

$$\frac{\gamma}{R_0} + \delta P(R) \cdot e^{(i\omega t - ikz)} - 2\mu\left(\frac{\partial v_r}{\partial r}\right)_{r=R} = \frac{\gamma}{R_0} + \frac{\gamma \varepsilon_0}{R_0}(k^2 R_0^2 - 1)e^{(i\omega t - ikz)} \rightarrow$$

$$\delta P(R) = \frac{\gamma \varepsilon_0}{R_0}(k^2 R_0^2 - 1) + 2\mu\left(\frac{\partial u_r}{\partial r}\right)_{r=R} e^{-(i\omega t - ikz)} \qquad . \qquad (A8)$$

Since the surface moves with the velocity of the particle at the surface, the third boundary condition (Eq. (24)) can be written as

$$\frac{\partial R}{\partial t} + \frac{\partial R}{\partial z}(v_z)_{r=R} = (v_r)_{r=R} \quad . \tag{A9}$$

Taking into account that $v_z$ and $v_r$ are of the order $\varepsilon_0 \omega R_0$ and $\partial R/\partial z \propto -\varepsilon_0 k R_0$ (Eq. (19)), I can drop the second term in the right-hand-side of Eq. (A.9) because it is $\sim(\varepsilon_0)^2$. Thus, this boundary condition for a small deformation of the cylinder, $\varepsilon_0 \ll 1$, is

$$\frac{\partial R}{\partial t} = (v_r)_{r=R_0} \quad . \tag{A10}$$

## Appendix B

Let us first show that $\Phi_1 = Aikr J_1(ikr)$ is a solution of Eq. (32). Introducing a new variable $\xi = ikr$, Eq. (32) reduces to the following form:

$$\frac{d^2\Phi_1}{d\xi^2} - \frac{1}{\xi}\frac{d\Phi_1}{d\xi} + \Phi_1 = 0 \quad . \tag{B1}$$

Nothing that $\Phi_1 = \xi \phi(\xi)$ yields

$$\xi \frac{d^2\phi}{d\xi^2} + 2\frac{d\phi}{d\xi} - \frac{d\phi}{d\xi} - \frac{\phi}{\xi} + \xi\phi = 0 \rightarrow$$

$$\frac{d^2\phi}{d\xi^2} + \frac{1}{\xi}\frac{d\phi}{d\xi} + \phi\left(1 - \frac{1}{\xi^2}\right) = 0 \quad . \tag{B2}$$

As one can see, Eq. (B2) is the Bessel function of the first order; that observation yields that $\phi = A \cdot J_1(\xi)$. Thus, I have shown that $\Phi_1 = Aikr J_1(ikr)$ is a solution of Eq. (32). Substituting $\tilde{k}$ for $k$ and $\tilde{A}$ for $A$ in $\Phi_1$, I obtain that $\Phi_2 = \tilde{A}i\tilde{k}r J_1(i\tilde{k}r)$ is a solution of Eq. (33).

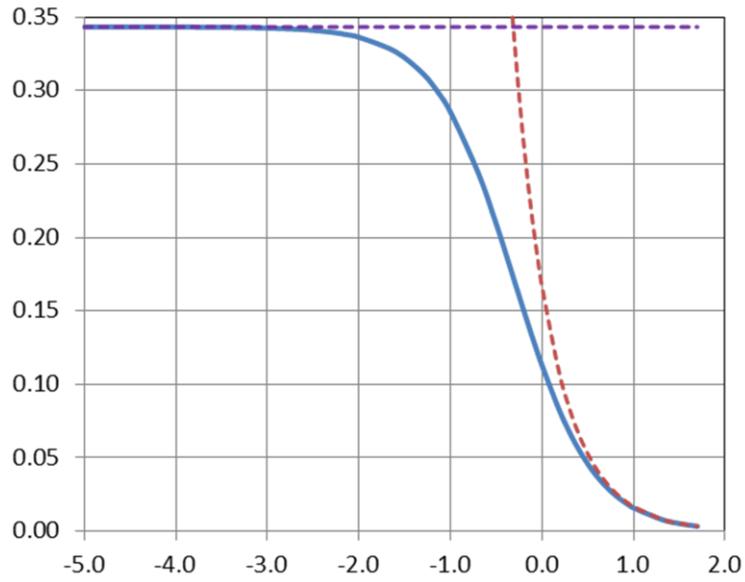

Fig. 1. Non-dimensional maximal increment of the Plateau-Rayleigh instability vs. $\log(\tilde{\mu})$ for a cylinder of a viscous liquid: the blue line is $-i\tilde{\omega}_{max}$; the red line is the Rayleigh asymptote for large $\tilde{\mu}$, Eq. (7); and the purple line is the Rayleigh asymptote for an inviscid liquid.

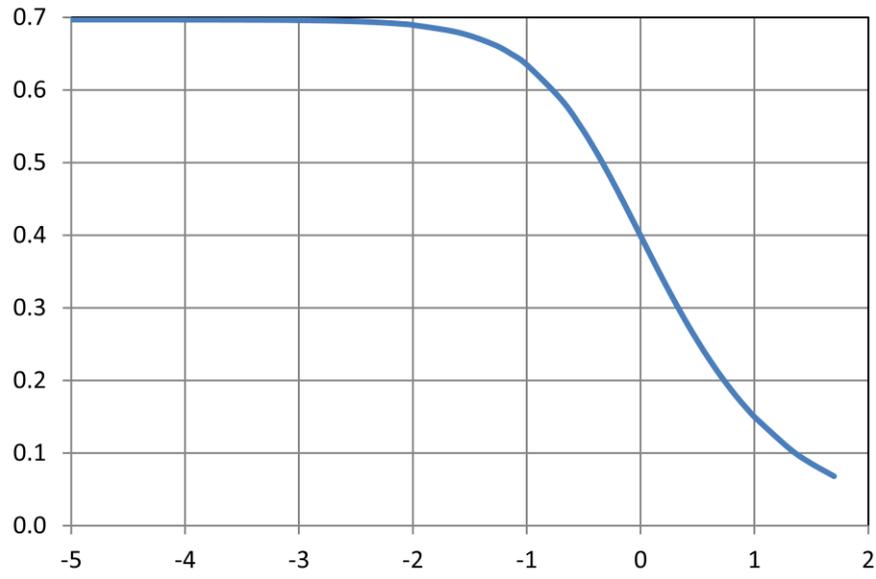

Fig. 2. $\xi_{max}$ vs. $\log(\tilde{\mu})$ that corresponds to $-i\tilde{\omega}_{max}$.

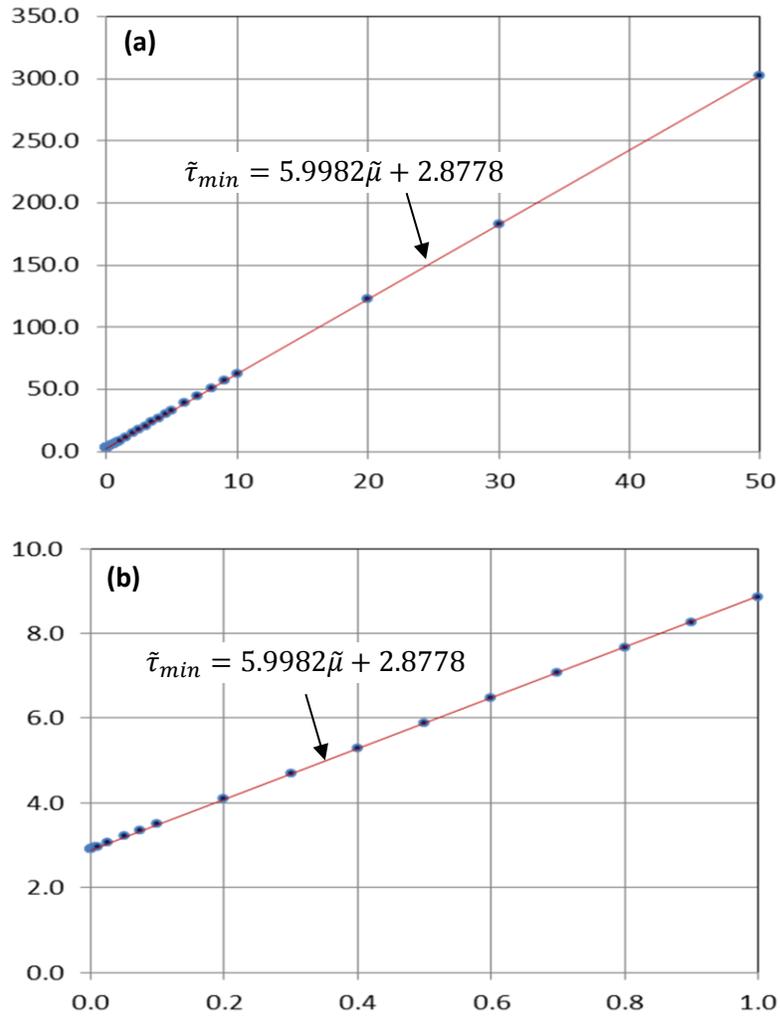

Fig. 3. $\tilde{\tau}_{min}$ vs. $\tilde{\mu}$: dots are $\tilde{\tau}_{min}$ calculated by the Mathematica code; the line is a linear approximation of $\tilde{\tau}_{min}$; (a) is the plot calculated in the full range of $\tilde{\mu}$; and (b) is the zoom of plot (a) for small $\tilde{\mu}$.

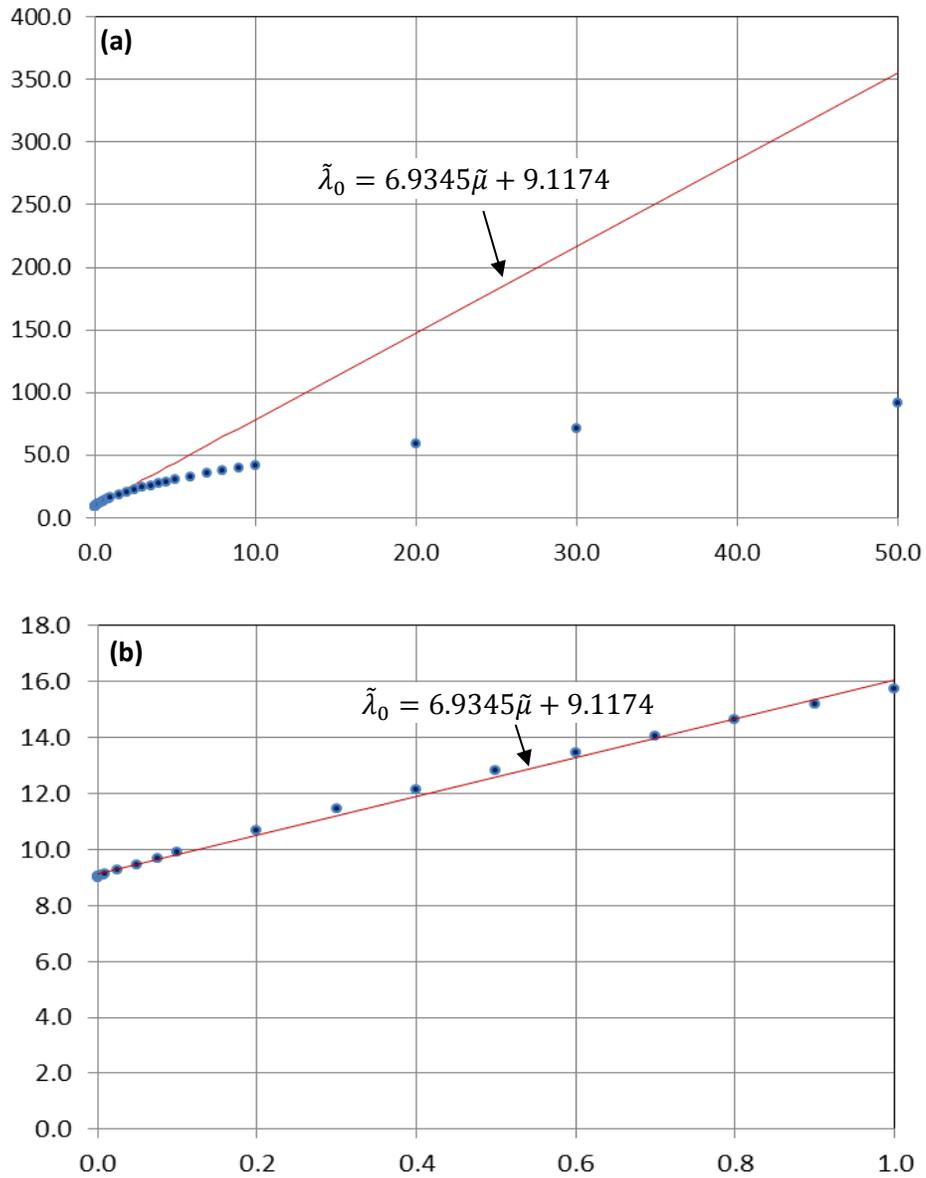

Fig. 4. $\tilde{\lambda}_0$ vs. $\tilde{\mu}$: dots are $\tilde{\lambda}_0$ calculated by the Mathematica code; the line is a linear approximation of $\tilde{\lambda}_0$ calculated for $0 \leq \tilde{\mu} \leq 1$; (a) is the plot calculated in the full range of $\tilde{\mu}$; and (b) is the zoom of plot (a) for small $\tilde{\mu}$.